\documentclass[11pt]{article}
\usepackage{txfonts}
\usepackage{color,epsfig}
\usepackage{mathrsfs}
\usepackage{amssymb}
\usepackage{amsfonts,graphicx,psfrag}

\newtheorem{Th}{Theorem}[section]

\newtheorem{lemma}{Lemma}[section]
\newtheorem{theorem}[lemma]{Theorem}

\newtheorem{proposition}[lemma]{Proposition}

\newtheorem{remark}[lemma]{Remark}
\let\lutzremark=\remark
\def\remark{\lutzremark\normalfont}

\newtheorem{notation}[lemma]{Notation}

\def\be{\begin{equation}}
\def\ee{\end{equation}}
\def\bea{\begin{eqnarray}}
\def\eea{\end{eqnarray}}
\def\bes{\begin{eqnarray*}}
\def\ees{\end{eqnarray*}}

\def\nn{\nonumber}
\def\<{\langle}
\def\>{\rangle}
\def\lb{\label}

\def\pt{\partial}

\def\R{{\bf R}}
\def\C{{\bf C}}
\def\Z{{\bf Z}}

\def\bb{{\beta}}

\def\th{{\theta}}

\def\om{{\omega}}

\def\diag{{\rm diag}}

\def\hb{\vrule height0.18cm width0.14cm $\,$}

\title{The symplectic reduction of the linearized Hamiltonian systems at elliptic relative equilibria of four-body problem}
\author{Qinglong Zhou$^{1} $\thanks{Partially supported by the Natural Science Foundation of Zhejiang Province (No.Y19A010072) and the Fundamental Research Funds for the Central Universities (No.2019QNA3002). E-mail: zhouqinglong@zju.edu.cn}\quad\\
	$^{1}$ Department of Mathematics\\Zhejiang University, Hangzhou 310027, Zhejiang, China}
\date{}

\begin{document}

\maketitle

\begin{abstract}
{In this paper, we consider the elliptic relative equilibria of four-body problem.
Here we
prove that the corresponding linearized Hamiltonian system at such an elliptic relative equilibria of $4$-bodies splits into two independent linear Hamiltonian systems, the first one is the
linearized Hamiltonian system of the Kepler $2$-body problem at Kepler elliptic orbit, and the
other system is the essential part of the linearized Hamiltonian system,
which is given implicitly. 
The reduction can be applied to the stability problem of such elliptic relative equilibria of four-body problem.}
\end{abstract}



\renewcommand{\theequation}{\thesection.\arabic{equation}}

\setcounter{equation}{0}
\setcounter{figure}{0}
\section{Introduction and main results}%{Section 1}
\label{sec:1}

For $n$ particles of mass $m_1,m_2,\ldots,m_n>0$,
let $q_1,q_2,\ldots,q_n\in \R^2$ the position vectors
respectively. Then the system of equations for $n$-body problem is
\be   m_i\ddot{q}_i=\frac{\partial U}{\partial q_i}, \qquad {\rm for}\quad i=1, 2, \ldots, n, \lb{1.1}\ee
where $U(q)=U(q_1,q_2,\ldots,q_n)=\sum_{1\leq i<j\leq n}\frac{m_im_j}{\|q_i-q_j\|}$ is the
potential or force function by using the standard norm $\|\cdot\|$ of vector in $\R^2$.

Note that $2\pi$-periodic solutions of this problem correspond to critical points of the action functional
$$ \mathcal{A}(q)=\int_{0}^{2\pi}\left[\sum_{i=1}^n\frac{m_i\|\dot{q}_i(t)\|^2}{2}+U(q(t))\right]dt $$
defined on the loop space $W^{1,2}(\R/2\pi\Z,\hat{\mathcal {X}})$, where
$$  \hat{\mathcal {X}}:=\left\{q=(q_1,q_2,\ldots,q_n)\in (\R^2)^n\,\,\left|\,\,
       \sum_{i=1}^n m_iq_i=0,\,\,q_i\neq q_j,\,\,\forall i\neq j \right. \right\}  $$
is the configuration space of the planar three-body problem.

Letting $p_i=m_i\dot{q}_i\in\R^2$ for $1\le i\le n$, then (\ref{1.1}) is transformed to a Hamiltonian system
\be \dot{p}_i=-\frac{\partial H}{\partial q_i},\,\,\dot{q}_i
  = \frac{\partial H}{\partial p_i},\qquad {\rm for}\quad i=1,2,\ldots, n,  \lb{1.2}\ee
with Hamiltonian function
\be H(p,q)=H(p_1,p_2,\ldots,p_n, q_1,q_2,\ldots,q_n)=\sum_{i=1}^n\frac{\|p_i\|^2}{2m_i}-U(q_1,q_2,\ldots,q_n).  \lb{1.3}\ee
A {\it central configuration} is a solution $(q_1,q_2,\ldots,q_n)=(a_1,a_2,\ldots,a_n)$ of
\begin{equation}
-\lambda m_iq_i={\pt U\over\pt q_i}(q_1,q_2,\ldots,q_n)
\end{equation}
for some constant $\lambda$.
An easy computation show that $\lambda={U(a)\over2I(a)}>0$,
where $I(a)={1\over2}\sum m_i||a_i||^2$ is the moment of inertia.
Please refer \cite{Win1} and \cite{Moe1} for the properties of central configuration.

It is well known that a planar central configuration of the $n$-body problem give rise to solutions
where each particle moves on a specific Keplerian orbit while the totaly of the particles move on a homographic motion.
Following Meyer and Schmidt \cite{MS}, we call these solutions as {\it elliptic relative equilibria}
and in shorthand notation, simply ERE.
Specially when $e=0$, the Keplerian elliptic
motion becomes circular motion and then all the bodies move around the center of masses along circular
orbits with the same frequency, which are called {\it relative equilibria} traditionally.

In the current paper, we given the precise reduction of the linearized Hamiltonian system at each elliptic relative equilibria of $4$-bodies.
To describe our main reduction result more precisely,
given positive masses $m=(m_1,m_2,m_3,m_4)\in(\R^+)^4$,
let $a=(a_1, a_2,a_3,a_4)$ be a central configuration of $m$ with
$a_i=(a_{ix},a_{iy})$ for $1\le i\le4$.
For convenience, we define four corresponding complex numbers:
\be
z_{a_i}=a_{ix}+\sqrt{-1}a_{iy},\quad i=1,2,3,4.  \label{complex.rep}
\ee

Without lose of generality, we normalize the three masses by
\begin{equation}\label{nomorlize.the.masses}
\sum_{i=1}^4 m_i=1,
\end{equation}
and normalize the positions $a_i,1\le i\le 4$ by
\bea
&&\sum_{i=1}^4 m_ia_i=0,  \label{mass.center}
\\
&&\sum_{i=1}^4 m_i|a_i|^2=2I(a)=1.\label{inertia}
\eea
Using the notations in (\ref{complex.rep}), (\ref{mass.center}) and (\ref{inertia}) are equivalent to
\bea
&&\sum_{i=1}^4 m_iz_{a_i}=0,  \label{mass.center'}
\\
&&\sum_{i=1}^4 m_i|z_{a_i}|^2=2I(a)=1.\label{inertia'}
\eea
Moreover, we define
\begin{equation}\label{mu}
\mu=U(a)=\sum_{1\le i<j\le 4}\frac{m_im_j}{|a_i-a_j|}=\sum_{1\le i<j\le 4}\frac{m_im_j}{|z_{a_i}-z_{a_j}|},
\quad
\sigma=(\mu p)^{1/4},
\end{equation}
and
\be\label{M}
\tilde{M}=\diag(m_1,m_2,m_3,m_4),\quad M=\diag(m_1,m_1,m_2,m_2,m_3,m_3,m_4,m_4).
\ee
%Because $a_1,a_2,a_3,a_4$ form a central configuration, we have
%\be\label{eq.of.cc}
%\sum_{j=1,j\ne i}^4\frac{m_j(z_{a_{j}}-z_{a_{i}})}{|z_{a_{i}}-z_{a_{j}}|^3}=
%\frac{U(a)}{2I(a)}z_{a_{i}}=\mu z_{a_{i}}.
%\ee

Let $B$ be a $4\times 4$ symmetric matrix such that
\begin{equation}\label{B}
B_{ij}=\left\{\begin{array}{c}
\frac{m_im_j}{|z_{a_i}-z_{a_j}|^3}\quad{if}\;i\ne j,1\le i,j\le 4,\\
-\sum_{j=1,j\ne i}^4\frac{m_im_j}{|z_{a_i}-z_{a_j}|^3}\quad{if}\;i=j,1\le i\le 4,
\end{array}\right.
\end{equation}
and
\bea
D&=&\mu I_4+\tilde{M}^{-1}B, \label{D}
\\
\tilde{D}&=&\mu I_4+\tilde{M}^{-1/2}B\tilde{M}^{-1/2}=\tilde{M}^{1/2}D\tilde{M}^{-1/2}. \label{tilde.D}
\eea
where $\mu$ is given by (\ref{mu}).

For convenience, we define two linear maps $\Phi,\Psi:\C\rightarrow \R^{2\times2}$ by
\bea
\Phi(z)=\left(\matrix{x & -y\cr
	y & x}\right) \quad\quad  
\forall z=x+\sqrt{-1}y\in\C,\;x,y\in\R,   \label{map1}
\\
\Psi(z)=\left(\matrix{x & y\cr
	y & -x}\right) \quad\quad  
\forall z=x+\sqrt{-1}y\in\C,\;x,y\in\R.   \label{map2}
\eea
Our main result is the following:

\begin{theorem}\label{main.theorem.decomposition}
In the planar $4$-body problem with given masses $m=(m_1,m_2,m_3,m_4)\in (\R^+)^4$, denote the
ERE with eccentricity $e\in [0,1)$ for $m$ by $q_{m,e}(t)=(q_1(t), q_2(t), q_3(t),q_4(t))$.
The linearized Hamiltonian system at $q_{m,e}$ is reduced to the summation of two independent
Hamiltonian systems, the first one is the linearized system of the Kepler $2$-body problem at the corresponding Kepler orbit,
the second one is the essential part of the linearized Hamiltonian system
which is given by
\begin{equation}\label{Nonconvex:LHS}
z'=J\left(\matrix{I& O& -J&  O\cr
	O& I&  O& -J\cr
	J& O& I_2-{r\over p}[{3\over2}I_2+\Psi(\beta_{11})]& 
	-{r\over p}\Psi(\beta_{12})\cr
	O& J& -{r\over p}\Psi(\beta_{12})& 
	I_2-{r\over p}[{3+\beta_2\over2}I_2+\Psi(\beta_{11})]
                }\right)z,
\end{equation}
where $\bb_2=1-{tr(D)\over\mu}$ 
and $\bb_{11},\bb_{12},\bb_{22}$ are given by (\ref{bb_11})-(\ref{bb_22}) below.
\end{theorem}

In Section 2 of this paper we focus on the proof of Theorem \ref{main.theorem.decomposition}.
In Appendix, we give some properties of $\Phi$ and $\Psi$.

\setcounter{equation}{0}%\setcounter{figure}{0}

\section{The symplectic reduction of the linearized Hamiltonian systems at  elliptic relative equilibria}\label{sec:2}

In \cite{MS} (cf. p.275), Meyer and Schmidt gave the essential part of the fundamental solution of the
elliptic Lagrangian orbit. Their method is explained in \cite{Lon5} too. Our study on ERE is based upon their method.

As in Section 1, 
for the given masses $m=(m_1,m_2,m_3,m_4)$ satisfying (\ref{nomorlize.the.masses}),
suppose the four particles are located at $a_1=(a_{1x},a_{1y}),a_2=(a_{2x},a_{2y}),a_3=(a_{3x},a_{3y}),a_4=(a_{4x},a_{4y})$,
form a central configuration.
%For convenience, we define four corresponding complex numbers:
%\be
%z_{a_i}=a_{ix}+\sqrt{-1}a_{iy},\quad i=1,2,3,4.  \label{complex.rep}
%\ee
%
%Without lose of generality, we normalize the three masses by
%\begin{equation}\label{nomorlize.the.masses}
%\sum_{i=1}^n m_i=1,
%\end{equation}
%and normalize the positions $a_i,1\le i\le 4$ by
%\bea
%&&\sum_{i=1}^4 m_ia_i=0,  \label{mass.center}
%\\
%&&\sum_{i=1}^4 m_i|a_i|^2=2I(a)=1.\label{inertia}
%\eea
%Using the notations in (\ref{complex.rep}), (\ref{mass.center}) and (\ref{inertia}) are equivalent to
%\bea
%&&\sum_{i=1}^4 m_iz_{a_i}=0,  \label{mass.center'}
%\\
%&&\sum_{i=1}^4 m_i|z_{a_i}|^2=2I(a)=1.\label{inertia'}
%\eea
%Moreover, we define
%\begin{equation}\label{mu}
%\mu=U(a)=\sum_{1\le i<j\le 4}\frac{m_im_j}{|a_i-a_j|}=\sum_{1\le i<j\le 4}\frac{m_im_j}{|z_{a_i}-z_{a_j}|},
%\quad
%\sigma=(\mu p)^{1/4},
%\end{equation}
%and
%\be
%\tilde{M}=\diag(m_1,m_2,m_3,m_4),\quad M=\diag(m_1,m_1,m_2,m_2,m_3,m_3,m_4,m_4).
%\ee
Using notations (\ref{complex.rep})-(\ref{M}),
since $a_1,a_2,a_3,a_4$ form a central configuration, we have
\be\label{eq.of.cc}
\sum_{j=1,j\ne i}^4\frac{m_j(z_{a_{j}}-z_{a_{i}})}{|z_{a_{i}}-z_{a_{j}}|^3}=
\frac{U(a)}{2I(a)}z_{a_{i}}=\mu z_{a_{i}}.
\ee

%Let $B$ be a $4\times 4$ symmetric matrix such that
%\begin{equation}
%B_{ij}=\left\{\begin{array}{c}
%               \frac{m_im_j}{|z_{a_i}-z_{a_j}|^3}\quad{if}\;i\ne j,1\le i,j\le 4,\\
%               -\sum_{j=1,j\ne i}^4\frac{m_im_j}{|z_{a_i}-z_{a_j}|^3}\quad{if}\;i=j,1\le i\le 4,
%              \end{array}\right.
%\end{equation}
%and
%\bea
%D&=&\mu I_4+\tilde{M}^{-1}B, \label{D}
%\\
%\tilde{D}&=&\mu I_4+\tilde{M}^{-1/2}B\tilde{M}^{-1/2}=\tilde{M}^{1/2}D\tilde{M}^{-1/2}. \label{tilde.D}
%\eea
%where $\mu$ is given by (\ref{mu}).

Based on matrix $B$ of (\ref{B}), 
$D$ has two simple eigenvalues: $\lambda_1=\mu$ with $v_1=(1,1,\ldots,1)^T$, and $\lambda_2=0$ with
$v_2=(z_{a_1},z_{a_2},z_{a_3},z_{a_4})^T$.
Exactly, we have
\bea
(Dv_1)_i&=&\mu-\sum_{j=1,j\ne i}^4\frac{m_j}{|z_{a_i}-z_{a_j}|^3}+\sum_{j=1,j\ne i}^4\frac{m_j}{|z_{a_i}-z_{a_j}|^3}=\mu,
\\
(Dv_2)_i&=&(\mu-\sum_{j=1,j\ne i}^4\frac{m_j}{|z_{a_i}-z_{a_j}|^3})z_{a_i}+\sum_{j=1,j\ne i}^4\frac{m_jz_{a_j}}{|z_{a_i}-z_{a_j}|^3}
\nonumber
\\
&=&\mu z_{a_i}+\sum_{j=1,j\ne i}^\frac{m_j(z_{a_j}-z_{a_i})}{|z_{a_i}-z_{a_j}|^3}
\nonumber
\\
&=&\mu z_{a_i}-\mu z_{a_i}
\nonumber
\\
&=&0,
\eea
where in the second last equality, we used (\ref{eq.of.cc}).
Moreover by (\ref{nomorlize.the.masses})-(\ref{inertia}), we have
\bea
\overline{v}_1^T\tilde{M}v_1&=&\sum_{i=1}^n m_i=1,\label{v1.M.v1}
\\
\overline{v}_1^T\tilde{M}v_2&=&\sum_{i=1}^4 m_iz_{a_i}=0,\label{v1.M.v2}
\\
\overline{v}_2^T\tilde{M}v_1&=&\sum_{i=1}^4 m_i\overline{z}_{a_i}=0,\label{v1.M.v2}
\\
\overline{v}_2^T\tilde{M}v_2&=&\sum_{i=1}^n m_i|z_{a_i}|^2=1.\label{v2.M.v2}
\eea
Let $\overline{v}_2=(\overline{z}_{a_1},\overline{z}_{a_2},\overline{z}_{a_3},\overline{z}_{a_4})^T$.
Because, $a_1,a_2,a_3,a_4$ forms a nonlinear central configuration,
$\overline{v}_2$ is independent with $v_2$. Moreover, $\overline{v}_2$ is also independent with $v_1$.
So $\overline{v}_2$ is another eigenvector of $D$ corresponding to eigenvalue $\lambda_3=0$.

Now, we construct $v_3$. We suppose
\be
v_3=k\overline{v}_2+lv_2  \label{v_3}
\ee
with $k\in\R,l\in\C$ will be given later.
If ${v}_2^T\tilde{M}v_2=\sum_{i=1}^n m_iz_{a_i}^2=0$, we set $k=1,l=0$, i.e., $v_3=\overline{v}_2$.
Then we have
\bea
\overline{v}_1^T\tilde{M}v_3&=&\sum_{i=1}^4 m_i\overline{z}_{a_i}=0,\label{v1.M.v3}
\\
\overline{v}_2^T\tilde{M}v_3&=&\sum_{i=1}^4 m_i\overline{z}_{a_i}^2=0,\label{v2.M.v3}
\\
\overline{v}_3^T\tilde{M}v_3&=&\sum_{i=1}^n m_i|z_{a_i}|^2=1.\label{v3.M.v3}
\eea
In the other cases, we also hope (\ref{v1.M.v3})-(\ref{v3.M.v3}) are satisfied.
Thus we have
\bea
0&=&\overline{v}_2^T\tilde{M}v_3=\overline{v}_2^T\tilde{M}(k\overline{v}_2+lv_2)=k\sum_{i=1}^4 m_i\overline{z}_{a_i}^2+l,
\\
1&=&\overline{v}_3^T\tilde{M}v_3=(kv_2+\overline{l}\overline{v}_2)^T\tilde{M}(k\overline{v}_2+lv_2)
=k^2+|l|^2+kl\sum_{i=1}^4 m_iz_{a_i}^2+k\overline{l}\sum_{i=1}^4 m_i\overline{z}_{a_i}^2.
\eea
Therefore, we have
\bea
k=\frac{1}{\sqrt{1-|\sum_{i=1}^4 m_i\overline{z}_{a_i}^2|^2}}, \label{param.k}
\\
l=-\frac{\sum_{i=1}^4 m_i\overline{z}_{a_i}^2}{\sqrt{1-|\sum_{i=1}^4 m_i\overline{z}_{a_i}^2|^2}}. \label{param.l}
\eea

We now construct a unitary matrix $\tilde{A}$ based on $v_1,v_2$ and $v_3$.
That is
\be
\tilde{A}=
\left(
\matrix{
1\quad z_{a_1}\quad b_1\quad c_1\cr
1\quad z_{a_2}\quad b_2\quad c_2\cr
1\quad z_{a_3}\quad b_3\quad c_3\cr
1\quad z_{a_4}\quad b_4\quad c_4
}
\right),
\ee
where $(b_1,b_2,b_3,b_4)=v_3^T$,i.e., $b_i=k\overline{z}_{a_i}+lz_{a_i},1\le i\le4$.
Then $c_i=A_{i4}$, where $A_{i4}$ is the algebraic cofactor of $c_i$.

In the other hand, the signed area of the triangle formed by $a_i,a_j$ and $a_k$ is given by
\be
\Delta_{ijk}=\frac{\sqrt{-1}}{4}\det
\left(
\matrix{
1\quad z_{a_1}\quad \overline{z}_{a_1}\cr
1\quad z_{a_2}\quad \overline{z}_{a_2}\cr
1\quad z_{a_3}\quad \overline{z}_{a_3}
}
\right).
\ee
Then $c_1=\overline{4k\sqrt{-1}\Delta_{234}}=-4k\sqrt{-1}\Delta_{234}$ and so on.
Note that, for any $\om\in\C,|\om|=1$, if $c_i$ are replaced by $\om c_i,i=1,2,3,4$, $\tilde{A}$ is also a unitary matrix.
Thus we can let
\be
(c_1,c_2,c_3,c_4)=({4k\rho\over m_1}\Delta_{234},-{4k\rho\over m_2}\Delta_{134},{4k\rho\over m_3}\Delta_{124},-{4k\rho\over m_4}\Delta_{123}), \label{c}
\ee
where
\be
\rho=\sqrt{m_1m_2m_3m_4}.
\ee
For convenience, we also write $v_4$ as
\be
v_4=(c_1,c_2,c_3,c_4)^T\in\R^4.  \label{v_4}
\ee
Now $v_1,v_2,v_3,v_4$ forms a unitary basis of $\C^n$.
Note that $v_1,v_2,v_3$ are eigenvectors of matrix $D$, then $v_4$ is also an eigenvector of $D$ with the corresponding eigenvalue
\be
\lambda_4=tr(D)-\lambda_1-\lambda_2-\lambda_3=tr(D)-\mu.
\ee
Moreover, we define
\bea
\bb_1&=&-\frac{\lambda_3}{\mu}=0,  \label{bb1}
\\
\bb_2&=&-\frac{\lambda_4}{\mu}=1-\frac{tr(D)}{\mu}.  \label{bb2}
\eea

In the following, if there is no confusion, we will use $a_i$ to represent $z_{a_i},1\le i\le4$.
By the definition of (\ref{v_3}) and (\ref{v_4}),
$Dv_k=\lambda_kv_k,k=3,4$ reads
\bea
\mu b_i-\sum_{j=1,j\ne i}^4\frac{m_j(b_j-b_i)}{|a_i-a_j|^3}=\lambda_3 b_i,\quad 1\le i\le 4,
\\
\mu c_i-\sum_{j=1,j\ne i}^4\frac{m_j(c_j-c_i)}{|a_i-a_j|^3}=\lambda_4 c_i,\quad 1\le i\le 4,
\eea
Let
\be
F_i=\sum_{j=1,j\ne i}^4\frac{m_im_j(b_i-b_j)}{|a_i-a_j|^3},\;
G_i=\sum_{j=1,j\ne i}^4\frac{m_im_j(c_i-c_j)}{|a_i-a_j|^3},\quad 1\le i\le 4, \label{F_ki}
\ee
then we have
\be
F_i=(\mu-\lambda_3)m_ib_i=\mu(1+\bb_1)m_ib_i,\quad G_i=(\mu-\lambda_4)m_ic_i=\mu(1+\bb_2)m_ic_i. \label{Fi.bi}
\ee
%Moreover, we have
%\be
%\sum_{i=1}^nF_{ik}b_{ik}=\sum_{i=1}^n(\mu-\lambda_k)m_ib_{ik}^2=\mu-\lambda_k=\mu(1+\bb_{k-2}), \label{sum.of.Fi.bi}
%\ee
%where in the last equality, we used (\ref{bi}).

%For later convenience, we introduce a new equation about $b_i,c_i,1\le i\le4$:
%\be
%\sum_{1\le i<j\le4}\frac{m_im_j(b_i-b_j)(c_i-c_j)}{|a_i-a_j|^3}=0. \label{eq.BC.8'}
%\ee
%By (\ref{AMA}), $\tilde{M}^{1/2}\tilde{A}$ is an orthogonal matrix, thus we have
%\be
%c_i=\frac{\rho}{m_i}\tilde{A}_{i4},
%\ee
%where $\tilde{A}_{i4}$ is the algebraic complement of $c_i$.
%Then we have
%\bea
%&&\sum_{1\le i<j\le4}\frac{m_im_j(b_i-b_j)(c_i-c_j)}{|a_i-a_j|^3}
%\nonumber
%\\
%&&=\rho\sum_{1\le i<j\le4}\frac{(b_i-b_j)(m_j\tilde{A}_{i4}-m_i\tilde{A}_{j4})}{|a_i-a_j|^3}
%\nonumber\\
%&&=\rho\left(\sum_{1\le i<j\le4}\tilde{A}_{i4}\frac{m_j(b_i-b_j)}{|a_i-a_j|^3}
%-\sum_{1\le i<j\le4}\tilde{A}_{j4}\frac{m_i(b_i-b_j)}{|a_i-a_j|^3}\right)
%\nonumber
%\\
%&&=\rho\left(\sum_{1\le i<j\le4}\tilde{A}_{i4}\frac{m_j(b_i-b_j)}{|a_i-a_j|^3}
%+\sum_{1\le j<i\le4}\tilde{A}_{i4}\frac{m_j(b_i-b_j)}{|a_i-a_j|^3}\right)
%\nonumber
%\\
%&&=\rho\sum_{i=1}^4\tilde{A}_{i4}\sum_{j=1,j\ne i}^4\frac{m_j(b_i-b_j)}{|a_i-a_j|^3}.
%\eea
%Let
%\begin{equation}
%F=
%\left(
%\matrix{
%1\quad a_{1x}\quad b_1\quad F_1/m_1\cr
%1\quad a_{2x}\quad b_2\quad F_2/m_2\cr
%1\quad a_{3x}\quad b_3\quad F_3/m_3\cr
%1\quad a_{4x}\quad b_4\quad F_4/m_4
%}
%\right).
%\end{equation}
%Then we obtain
%\be
%\sum_{1\le i<j\le4}\frac{m_im_j(b_i-b_j)(c_i-c_j)}{|a_i-a_j|^3}=\det F=0,
%\ee
%where the last equality is obtained by (\ref{Fi.bi}).

Now as in p.263 of \cite{MS}, Section 11.2 of \cite{Lon5}, we define
\begin{equation}\label{PQYX}
P=\left(\matrix{p_1\cr p_2\cr p_3\cr p_4}\right),
\quad
Q=\left(\matrix{q_1\cr q_2\cr q_3\cr q_4}\right),
\quad
Y=\left(\matrix{G\cr Z\cr W_1\cr  W_2}\right),
\quad
X=\left(\matrix{g\cr z\cr w_1\cr w_2}\right),
\end{equation}
where $p_i$, $q_i$, $i=1,2,3,4$ and $G$, $Z$, $W_1\;W_2$, $g$, $z$, $w_1,w_2$ are all column vectors in $\R^2$.
We make the symplectic coordinate change
\be\lb{transform1}  P=A^{-T}Y,\quad Q=AX,  \ee
where the matrix $A$ is constructed as in the proof of Proposition 2.1 in \cite{MS}.
Concretely, the matrix $A\in {\bf GL}(\R^{8})$ is given by
\begin{equation}
A=
\left(
\matrix{
I\quad A_1\quad B_{1}\quad C_{1}\cr
I\quad A_2\quad B_{2}\quad C_{2}\cr
I\quad A_3\quad B_3\quad   C_3\cr
I\quad A_4\quad B_{4}\quad C_4
}
\right),
\end{equation}
where each $A_i$ is a $2\times2$ matrix given by
\begin{eqnarray}
A_i &=& (a_i, Ja_i)=\Phi(a_i),  \label{Aa}
\\
B_i &=& (b_i, Jb_i)=\Phi(b_i),  \label{Bb}
\\
C_i &=& (c_i, Jc_i)=\Phi(c_i)=c_iI_2, \label{Cc}
\end{eqnarray}
where $\Phi$ is given by (\ref{map1}).
Moreover, by the definition of $v_i,1\le i\le 4$, we obtain
\bea
\overline{\tilde{A}}^T\tilde{M}\tilde{A}&=&(\overline{v}_1,\overline{v}_2,\overline{v}_3,\overline{v}_4)^T\tilde{M}(v_1,v_2,v_3,v_4)
=I_4
\eea
By (\ref{map.matrix}), we have
$A^TMA=\Phi(\tilde{A})^T\Phi(\tilde{M})\Phi(\tilde{A})=\Phi(\overline{\tilde{A}}^T\tilde{M}\tilde{A})=\Phi(I_4)=I_{8}$
is fulfilled (cf. (13) in p.263 of \cite{MS}).

Now we consider the Hamiltonian function of the four-body problem.
Under the coordinate change (\ref{transform1}), we get the kinetic enrgy
\begin{equation}
K=\frac{1}{2}(|G|^2+|Z|^2+|W_1|^2+|W_2|^2),
\end{equation}
and the potential function
\begin{eqnarray}
U_{ij}(z,w_1,w_2)&=&\frac{m_im_j}{d_{ij}(z,w_1,w_2)}, \label{U_ij}
\\
U(z,w_1,w_2)&=&\sum_{1\le i<j\le 4}U_{ij}(z,w_1,w_2),\label{U}
\end{eqnarray}
with
\begin{eqnarray}
d_{ij}(z,w_1,w_2)&=&|(A_i-A_j)z+(B_i-B_j)w_1+(C_i-C_j)w_2|
\nonumber
\\
&=&|\Phi(a_i-a_j)z+\Phi(b_i-b_j)w_1+\Phi(c_i-c_j)w_2|,
\end{eqnarray}
where we used (\ref{Aa})-(\ref{Cc}).

Let $\theta$ be the true anomaly.
Then under the same steps of symplectic transformation in the proof of Theorem 11.10 (p. 100 of \cite{Lon5}),
the resulting Hamiltonian function of the 3-body problem is given by
\begin{eqnarray}\label{new.H.function}
&&H(\theta,\bar{Z},\bar{W_1},\bar{W}_2,\bar{z},\bar{w_1},\bar{w}_2)=\frac{1}{2}(|\bar{Z}|^2+|\bar{W_1}|^2+|\bar{W_2}|^2)
+(\bar{z}\cdot J\bar{Z}+\bar{w_1}\cdot J\bar{W_1}+\bar{w_2}\cdot J\bar{W_2})
\nonumber
\\
&&\quad\quad\quad\quad\quad+\frac{p-r(\theta)}{2p}(|\bar{z}|^2+|\bar{w_1}|^2+|\bar{w_2}|^2)
-\frac{r(\theta)}{\sigma}U(\bar{z},\bar{w_1},\bar{w}_2),
\end{eqnarray}
where $\mu$ is given by (\ref{mu}) and
\begin{equation}
r(\theta)=\frac{p}{1+e\cos\theta}.
\end{equation}

We now derived the linearized Hamiltonian system at the elliptic relative equilibrium.
\begin{proposition}\label{linearized.Hamiltonian}
Using notations in (\ref{PQYX}), elliptic relative equilibria $(P(t),Q(t))^T$ of the system (\ref{1.2}) with
\begin{equation}
Q(t)=(r(t)R(\theta(t))a_1,r(t)R(\theta(t))a_2,r(t)R(\theta(t))a_3,r(t)R(\theta(t))a_4)^T,\quad P(t)=M\dot{Q}(t)
\end{equation}
in time $t$ with the matrix $M=diag(m_1,m_1,m_2,m_2,m_3,m_3,m_4,m_4)$,
is transformed to the new solution $(Y(\theta),X(\theta))^T$ in the variable true anomaly $\theta$
with $G=g=0$ with respect to the original Hamiltonian function $H$ of (\ref{new.H.function}), which is given by
\begin{equation}
Y(\theta)=\left(
\matrix{
\bar{Z}(\theta)\cr
\bar{W}_1(\theta)\cr
\bar{W}_2(\theta)}
\right)
=\left(
\matrix{
0\cr
\sigma\cr
0\cr
0\cr
0\cr
0}
\right),
\quad
X(\theta)=\left(
\matrix{
\bar{z}(\theta)\cr
\bar{w_1}(\theta)\cr
\bar{w_2}(\theta)}
\right)
=\left(
\matrix{
\sigma\cr
0\cr
0\cr
0\cr
0\cr
0}
\right).
\end{equation}

Moreover, the linearized Hamiltonian system at the elliptic relative equilibria
${\xi}_0\equiv(Y(\theta),X(\theta))^T =$
\newline
$(0,\sigma,0,0,0,0,\sigma,0,0,0,0,0)^T\in\R^{12}$
depending on the true anomaly $\theta$ with respect to the Hamiltonian function
$H$ of (\ref{new.H.function}) is given by
\begin{equation}
\dot\zeta(\theta)=JB(\theta)\zeta(\theta),  \label{general.linearized.Hamiltonian.system}
\end{equation}
with
\begin{eqnarray}
B(\theta)&=&H''(\theta,\bar{Z},\bar{W_1},\bar{W}_2,\bar{z},\bar{w_1},\bar{w}_2)|_{\bar\xi=\xi_0}
\nonumber
\\
&=&\left(
\matrix{
I& O& O& -J&  O&  O\cr
O& I& O&  O& -J&  O\cr
O& O& I&  O&  O& -J\cr
J& O& O& H_{\bar{z}\bar{z}}(\theta,\xi_0)& O& O\cr
O& J& O& O& H_{\bar{w_1}\bar{w_1}}(\theta,\xi_0)& H_{\bar{w_1}\bar{w_2}}(\theta,\xi_0)\cr
O& O& J& O& H_{\bar{w_2}\bar{w_1}}(\theta,\xi_0)& H_{\bar{w_2}\bar{w_2}}(\theta,\xi_0)
}
\right),
\end{eqnarray}
and
\begin{eqnarray}
H_{\bar{z}\bar{z}}(\theta,\xi_0)&=&\left(
\matrix{
-\frac{2-e\cos\theta}{1+e\cos\theta} & 0\cr
0 & 1
}
\right),
\quad
\\
H_{\bar{w_i}\bar{w_i}}(\theta,\xi_0)&=&I_2-\frac{r}{p}\left[\frac{3+\bb_i}{2}I_2+\Psi(\bb_{ii})\right],\ \ i=1,2, \label{H_12}
\\
H_{\bar{w_1}\bar{w_2}}(\theta,\xi_0)&=&-\frac{r}{p}\Psi(\bb_{12}),
\end{eqnarray}
where $\bb_1=0$ and
$\bb_2$ are given by (\ref{bb2}),
and $\bb_{11},\bb_{12},\bb_{22}$ are given by
\bea
\bb_{11}&=&{3\over2\mu}\sum_{1\le i<j\le 4}\frac{m_im_j(a_i-a_j)^2(\overline{b}_i-\overline{b}_j)^2}{|a_i-a_j|^5},  \label{bb_11}
\\
\bb_{12}&=&{3\over2\mu}\sum_{1\le i<j\le 4}\frac{m_im_j(a_i-a_j)^2(\overline{b}_i-\overline{b}_j)(\overline{c}_i-\overline{c}_j)}{|a_i-a_j|^5},
\label{bb_12}
\\
\bb_{22}&=&{3\over2\mu}\sum_{1\le i<j\le 4}\frac{m_im_j(a_i-a_j)^2(\overline{c}_i-\overline{c}_j)^2}{|a_i-a_j|^5},\label{bb_22}
\eea
and $H''$ is the Hession Matrix of $H$ with respect to its variable $\bar{Z}$,
$\bar{W_1},\bar{W}_2$, $\bar{z}$, $\bar{w_1},\bar{w}_2$.
The corresponding quadratic Hamiltonian function is given by
\begin{eqnarray}
H_2(\theta,\bar{Z},\bar{W_1},\bar{W}_2,\bar{z},\bar{w_1},\bar{w}_2)
&=&\frac{1}{2}|\bar{Z}|^2+\bar{Z}\cdot J\bar{z}+\frac{1}{2}H_{\bar{z}\bar{z}}(\theta,\xi_0)|\bar{z}|^2+H_{\bar{w_1}\bar{w_2}}(\theta,\xi_0)\bar{w_1}\cdot\bar{w_2}
\nonumber\\
&&+\left(\frac{1}{2}|\bar{W_1}|^2+\bar{W_1}\cdot J\bar{w_1}+\frac{1}{2}H_{\bar{w_1}\bar{w_1}}(\theta,\xi_0)|\bar{w_1}|^2\right)
\nonumber\\
&&+\left(\frac{1}{2}|\bar{W_2}|^2+\bar{W_2}\cdot J\bar{w_2}+\frac{1}{2}H_{\bar{w_2}\bar{w_2}}(\theta,\xi_0)|\bar{w_2}|^2\right).
\end{eqnarray}
\end{proposition}

{\bf Proof.} The proof is similar to those of Proposition 11.11 and Proposition 11.13 of \cite{Lon5}.
We just need to compute $H_{\bar{z}\bar{z}}(\theta,\xi_0)$, $H_{\bar{z}\bar{w_i}}(\theta,\xi_0)$
and $H_{\bar{w_i}\bar{w_j}}(\theta,\xi_0)$ for $i,j=1,2$.

For simplicity, we omit all the upper bars on the variables of $H$ in (\ref{new.H.function}) in this proof.
By (\ref{new.H.function}), we have
\bea
H_z&=&JZ+\frac{p-r}{p}z-\frac{r}{\sigma}U_z(z,w_1,w_2),  \nn\\
H_{w_i}&=&JW_i+\frac{p-r}{p}w_i-\frac{r}{\sigma}U_{w_i}(z,w_1,w_2), \quad i=1,2, \nn
\eea
and
\be\lb{Hessian}\left\{
\begin{array}{l}
H_{zz}=\frac{p-r}{p}I-\frac{r}{\sigma}U_{zz}(z,w_1,w_2),
\\
H_{zw_i}=H_{w_lz}=-\frac{r}{\sigma}U_{zw_i}(z,w_1,w_2),\quad i=1,2,
\\
H_{w_iw_i}=\frac{p-r}{p}I-\frac{r}{\sigma}U_{w_iw_i}(z,w_1,w_2),\quad i=1,2,
\\
H_{w_1w_2}=H_{w_2w_1}=-\frac{r}{\sigma}U_{w_1w_2}(z,w_1,w_2),
\end{array}\right. \ee
where we write $H_z$ and $H_{zw_i}$ etc to denote the derivative of $H$ with respect to $z$,
and the second derivative of $H$ with respect to $z$ and then $w_i$ respectively.
Note that all the items above are $2\times2$ matrices.

For $U_{ij}$ defined in (\ref{U_ij}) with $1\le i<j\le n,1\le l\le n-2$,
we have
\bea
\frac{\partial U_{ij}}{\partial z}(z,w_1,w_2) &=& -\frac{m_im_j\Phi(a_i-a_j)^T}{|\Phi(a_i-a_j)z+\Phi(b_i-b_j)w_1+\Phi(c_i-c_j)w_2|^3}
\nn\\
&&\qquad\cdot\left[\Phi(a_i-a_j)z+\Phi(b_i-b_j)w_1+\Phi(c_i-c_j)w_2\right],
\\
\frac{\partial U_{ij}}{\partial w_1}(z,w_1,w_2) &=& -\frac{m_im_j\Phi(b_i-b_j)^T}{|\Phi(a_i-a_j)z+\Phi(b_i-b_j)w_1+\Phi(c_i-c_j)w_2|^3}
\nn\\
&&\qquad\cdot\left[\Phi(a_i-a_j)z+\Phi(b_i-b_j)w_1+\Phi(c_i-c_j)w_2\right],
\\
\frac{\partial U_{ij}}{\partial w_2}(z,w_1,w_2)&=& -\frac{m_im_j\Phi(c_i-c_j)^T}{|\Phi(a_i-a_j)z+\Phi(b_i-b_j)w_1+\Phi(c_i-c_j)w_2|^3}
\nn\\
&&\qquad\cdot\left[\Phi(a_i-a_j)z+\Phi(b_i-b_j)w_1+\Phi(c_i-c_j)w_2\right],
\eea
and
\bea
\frac{\partial^2 U_{ij}}{\partial z^2}(z,w_1,w_2)
&=&-\frac{m_im_j|a_i-a_j|^2I_2}{|\Phi(a_i-a_j)z+\Phi(b_i-b_j)w_1+\Phi(c_i-c_j)w_2|^3}  \nn\\
&&+3\frac{m_im_j}{|\Phi(a_i-a_j)z+\Phi(b_i-b_j)w_1+\Phi(c_i-c_j)w_2|^5}
\nn\\
&&\qquad\cdot\Phi(a_i-a_j)^T\left[\Phi(a_i-a_j)z+\Phi(b_i-b_j)w_1+\Phi(c_i-c_j)w_2\right]  \nn\\
&&\qquad\cdot\left[\Phi(a_i-a_j)z+\Phi(b_i-b_j)w_1+\Phi(c_i-c_j)w_2\right]^T\Phi(a_i-a_j),  \\
\frac{\partial^2 U_{ij}}{\partial z\partial w_1}(z,w_1,w_2)&=&
-\frac{m_im_j\Phi(a_i-a_j)^T\Phi(b_i-b_j)}{|\Phi(a_i-a_j)z+\Phi(b_i-b_j)w_1+\Phi(c_i-c_j)w_2|^3} \nn\\
&&+3\frac{m_im_j}{|\Phi(a_i-a_j)z+\Phi(b_i-b_j)w_1+\Phi(c_i-c_j)w_2|^5}
\nn\\
&&\qquad\cdot\Phi(a_i-a_j)^T\left[\Phi(a_i-a_j)z+\Phi(b_i-b_j)w_1+\Phi(c_i-c_j)w_2\right]  \nn\\
&&\qquad\cdot\left[\Phi(a_i-a_j)z+\Phi(b_i-b_j)w_1+\Phi(c_i-c_j)w_2\right]^T\Phi(b_i-b_j),  \\
\frac{\partial^2 U_{ij}}{\partial {w_1}^2}(z,w_1,w_2)
&=&-\frac{m_im_j|b_i-b_j|^2I_2}{|\Phi(a_i-a_j)z+\Phi(b_i-b_j)w_1+\Phi(c_i-c_j)w_2|^3}  \nn\\
&&+3\frac{m_im_j}{|\Phi(a_i-a_j)z+\Phi(b_i-b_j)w_1+\Phi(c_i-c_j)w_2|^5}
\nn\\
&&\qquad\cdot\Phi(b_i-b_j)^T\left[\Phi(a_i-a_j)z+\Phi(b_i-b_j)w_1+\Phi(c_i-c_j)w_2\right]   \nn\\
&&\qquad\cdot\left[\Phi(a_i-a_j)z+\Phi(b_i-b_j)w_1+\Phi(c_i-c_j)w_2\right]^T\Phi(b_i-b_j).
\eea

Let
$$ K=\left(\matrix{2 & 0\cr
                   0 & -1}\right), \quad
K_1=\left(\matrix{1 & 0\cr
                  0 & 0}\right),\quad
K_2=\left(\matrix{1 & 0\cr
                  0 & -1}\right)=\Psi(1),  $$
where $\Psi$ is given by (\ref{map2}).
Now evaluating these functions at the solution $\bar\xi_0=(0,\sigma,0,0,0,0,\sigma,0,0,0,0,0)^T\in\R^8$
 with $z=(\sigma,0)^T,w_i=(0,0)^T,1\le i\le 2$, and summing them up,
we obtain
\begin{eqnarray}
\frac{\partial^2 U}{\partial z^2}\left|_{\xi_0}\right.&=&
\sum_{1\le i<j\le 4}\frac{\partial^2 U_{ij}}{\partial z^2}\left|_{\xi_0}\right.
\nonumber\\
&=&\sum_{1\le i<j\le 4}\left(-\frac{m_im_j|a_i-a_j|^2}{|(a_i-a_j)\sigma|^3}I
                        +3\frac{m_im_j\sigma^2|a_i-a_j|^2K_1|a_i-a_j|^2}{|(a_i-a_j)\sigma|^5}\right)
\nonumber\\
&=&\frac{1}{\sigma^3}\left(\sum_{1\le i<j\le4}\frac{m_im_j}{|a_i-a_j|}\right)K
\nonumber\\
&=&\frac{\mu}{\sigma^3}K,  \label{U_zz}
\\
\frac{\partial^2 U}{\partial w_1^2}\left|_{\xi_0}\right.&=&
\sum_{1\le i<j\le 4}\frac{\partial^2 U_{ij}}{\partial w_l^2}\left|_{\xi_0}\right.
\nonumber\\
&=&\sum_{1\le i<j\le 4}\left(-\frac{m_im_j|b_i-b_j|^2}{|(a_i-a_j)\sigma|^3}I
                        +3\frac{m_im_j\sigma^2\Phi(b_i-b_j)^T\Phi(a_i-a_j)K_1\Phi(a_i-a_j)^T\Phi(b_i-b_j)}{|(a_i-a_j)\sigma|^5}\right)
\nonumber\\
&=&\sum_{1\le i<j\le 4}\left(-\frac{m_im_j|b_i-b_j|^2}{|(a_i-a_j)\sigma|^3}I
                        +3\frac{m_im_j\sigma^2\Phi(b_i-b_j)^T\Phi(a_i-a_j)\frac{I_2+K_2}{2}\Phi(a_i-a_j)^T\Phi(b_i-b_j)}{|(a_i-a_j)\sigma|^5}\right)
\nonumber\\
&=&\sum_{1\le i<j\le 4}\left(-\frac{m_im_j|b_i-b_j|^2}{|(a_i-a_j)\sigma|^3}I
                        +{3\over2}\frac{m_im_j\sigma^2\Phi(b_i-b_j)^T\Phi(a_i-a_j)\Phi(a_i-a_j)^T\Phi(b_i-b_j)}{|(a_i-a_j)\sigma|^5}\right)
\nonumber\\
&&+\sum_{1\le i<j\le 4}
\left({3\over2}\frac{m_im_j\sigma^2\Phi(b_i-b_j)^T\Phi(a_i-a_j)\Psi(1)\Phi(a_i-a_j)^T\Phi(b_i-b_j)}{|(a_i-a_j)\sigma|^5}\right)
\nonumber\\
&=&\sum_{1\le i<j\le 4}\left(-\frac{m_im_j|b_i-b_j|^2}{|(a_i-a_j)\sigma|^3}I
                        +{3\over2}\frac{m_im_j\sigma^2\Phi(|b_i-b_j|^2|a_i-a_j|^2)}{|(a_i-a_j)\sigma|^5}\right)
\nonumber\\
&&+\sum_{1\le i<j\le 4}
\left({3\over2}\frac{m_im_j\sigma^2\Psi((a_i-a_j)^2(\overline{b}_i-\overline{b}_j)^2)}{|(a_i-a_j)\sigma|^5}\right)
\nonumber\\
&=&{1\over2\sigma^3}\sum_{1\le i<j\le 4}\left(\frac{m_im_j|b_i-b_j|^2}{|a_i-a_j|^3}\right)I_2
+{1\over\sigma^3}\Psi\left({3\over2}\sum_{1\le i<j\le 4}
\frac{m_im_j(a_i-a_j)^2(\overline{b}_i-\overline{b}_j)^2}{|a_i-a_j|^5}\right)
\nonumber\\
&=&\frac{1}{2\sigma^3}\left(\sum_{i=1}^4 \overline{b}_i\sum_{j=1,j\ne i}^4\frac{m_im_j(b_i-b_j)}{|a_i-a_j|^3}\right)I_2
+{1\over\sigma^3}\Psi\left({3\over2}\sum_{1\le i<j\le 4}
\frac{m_im_j(a_i-a_j)^2(\overline{b}_i-\overline{b}_j)^2}{|a_i-a_j|^5}\right)
\nonumber
\\
&=&\frac{1}{2\sigma^3}\left(\sum_{i=1}^4 \overline{b}_iF_i\right)I_2
+{1\over\sigma^3}\Psi\left({3\over2}\sum_{1\le i<j\le 4}
\frac{m_im_j(a_i-a_j)^2(\overline{b}_i-\overline{b}_j)^2}{|a_i-a_j|^5}\right)
\nonumber\\
&=&\frac{\mu(1+\bb_1)}{2\sigma^3}I_2
+{1\over\sigma^3}\Psi\left({3\over2}\sum_{1\le i<j\le 4}
\frac{m_im_j(a_i-a_j)^2(\overline{b}_i-\overline{b}_j)^2}{|a_i-a_j|^5}\right)
\nonumber\\
&=&\frac{\mu(1+\bb_1)}{2\sigma^3}I_2
+{\mu\over\sigma^3}\Psi(\bb_{11}), \label{U_w1w1}
\end{eqnarray}
where in the third equality of the first formula, we used (\ref{F_ki}),
and in the last equality of the second formula, we use the definition (\ref{Fi.bi}) and (\ref{bb_11}).
Similarly, we have
\bea
\frac{\partial^2 U}{\partial w_2^2}\left|_{\xi_0}\right.&=&
\frac{\mu(1+\bb_2)}{2\sigma^3}I_2
+{1\over\sigma^3}\Psi\left({3\over2}\sum_{1\le i<j\le 4}
\frac{m_im_j(a_i-a_j)^2(\overline{c}_i-\overline{c}_j)^2}{|a_i-a_j|^5}\right)
\nonumber
\\
&=&\frac{\mu(1+\bb_2)}{2\sigma^3}I_2
+{\mu\over\sigma^3}\Psi(\bb_{22}), \label{U_w2w2}
\\
\frac{\partial^2 U}{\partial w_1\partial w_2}\left|_{\xi_0}\right.&=&
{1\over\sigma^3}\Psi\left({3\over2}\sum_{1\le i<j\le 4}
\frac{m_im_j(a_i-a_j)^2(\overline{b}_i-\overline{b}_j)(\overline{c}_i-\overline{c}_j)}{|a_i-a_j|^5}\right)
\nonumber\\
&=&{\mu\over\sigma^3}\Psi(\bb_{12}). \label{U_w1w2}
\eea

Moreover, we have
\begin{eqnarray}
\frac{\partial^2 U}{\partial z\partial w_1}\left|_{\xi_0}\right.&=&
\sum_{1\le i<j\le 4}\frac{\partial^2 U_{ij}}{\partial z\partial w_1}\left|_{\xi_0}\right.
\nonumber\\
&=&\sum_{1\le i<j\le4}\left(-\frac{m_im_j\Phi(a_i-a_j)^T\Phi(b_i-b_j)}{|(a_i-a_j)\sigma|^3}
                        +3\frac{m_im_j\sigma^2|a_i-a_j|^2K_1\Phi(a_i-a_j)^T\Phi(b_i-b_j)}{|(a_i-a_j)\sigma|^5}\right)
\nonumber\\
&=&\frac{K}{\sigma^3}\left(\sum_{1\le i<j\le 4}\frac{m_im_j\Phi((\overline{a}_i-\overline{a}_j)(b_i-b_j))}{|a_i-a_j|^3}\right)
\nonumber\\
&=&\frac{K}{\sigma^3}\Phi\left(\sum_{1\le i<j\le 4}\frac{m_im_j(\overline{a}_i-\overline{a}_j)(b_i-b_j)}{|a_i-a_j|^3}\right)
\nonumber\\
&=&\frac{K}{\sigma^3}\Phi\left(\sum_{1\le i<j\le 4}\frac{m_im_j\overline{a}_i(b_i-b_j)}{|a_i-a_j|^3}
-\sum_{1\le i<j\le 4}\frac{m_im_j\overline{a}_j(b_i-b_j)}{|a_i-a_j)|^3}\right)
\nonumber\\
&=&\frac{K}{\sigma^3}\Phi\left(\sum_{1\le i<j\le 4}\frac{m_im_j\overline{a}_i(b_i-b_j)}{|a_i-a_j|^3}
-\sum_{1\le j<i\le 4}\frac{m_jm_i\overline{a}_i(b_j-b_i)}{|a_j-a_i)|^3}\right)
\nonumber\\
&=&\frac{K}{\sigma^3}\Phi\left(\sum_{i=1}^4\overline{a}_i\sum_{j=i+1}^4\frac{m_im_j(b_i-b_j)}{|a_i-a_j|^3}
+\sum_{i=1}^4\overline{a}_i\sum_{j=1}^{i-1}\frac{m_im_j(b_i-b_j)}{|a_i-a_j|^3}\right)
\nonumber\\
&=&\frac{K}{\sigma^3}\Phi\left(\sum_{i=1}^4\overline{a}_i\sum_{j=1,j\ne i}^4\frac{m_im_j(b_i-b_j)}{|a_i-a_j|^3}\right)
\nonumber\\
&=&\frac{K}{\sigma^3}\Phi\left(\sum_{i=1}^4\overline{a}_iF_i\right)
\nonumber\\
&=&\frac{K}{\sigma^3}\Phi\left(\mu(1+\bb_1)\sum_{i=1}^4m_i\overline{a}_ib_i\right)
\nonumber\\
&=&O,\label{U_zw1}
\end{eqnarray}
where in the second last equation, we used (\ref{eq.of.cc}), and in the last equality, we used (\ref{Fi.bi}).
Similarly, we have
\be
\frac{\partial^2 U}{\partial z\partial w_2}\left|_{\xi_0}\right.=0. \label{U_zw2}
\ee

By \label{U_zw1} and (\ref{U_zz})-(\ref{U_zw2}), we have
\begin{eqnarray}
H_{zz}|_{\xi_0}&=&\frac{p-r}{p}I-\frac{r\mu}{\sigma^4}K=I-\frac{r}{p}I-\frac{r\mu}{p\mu}K
=I-\frac{r}{p}(I+K)
=\left(\matrix{-\frac{2-e\cos\theta}{1+e\cos\theta} & 0\cr
               0 & 1}\right),  \nn\\
H_{zw_i}|_{\xi_0}&=&-\frac{r}{\sigma}\frac{\partial^2U}{\partial z\partial w_i}|_{\xi_0}=O,\quad 1\le i\le 2,
\nn\\
H_{w_1w_1}|_{\xi_0}&=&\frac{p-r}{p}I-\frac{r}{\sigma}
   \left[\frac{\mu(1+\bb_1)}{2\sigma^3}I_2+{\mu\over\sigma^3}\Psi(\bb_{11})\right]
=I-\frac{r}{p}I-\frac{r}{p}\left[\frac{1+\bb_1}{2}I_2+\Psi(\bb_{11})\right]
\nonumber
\\
&=&I-\frac{r}{p}\left[\frac{3+\bb_1}{2}I_2+\Psi(\bb_{11})\right],
\nn\\
H_{w_2w_2}|_{\xi_0}&=&\frac{p-r}{p}I-\frac{r}{\sigma}
   \left[\frac{\mu(1+\bb_2)}{2\sigma^3}I_2+{\mu\over\sigma^3}\Psi(\bb_{22})\right]
=I-\frac{r}{p}I-\frac{r}{p}\left[\frac{1+\bb_2}{2}I_2+\Psi(\bb_{22})\right]
\nonumber
\\
&=&I-\frac{r}{p}\left[\frac{3+\bb_2}{2}I_2+\Psi(\bb_{22})\right],
\nn\\
H_{w_1w_2}|_{\xi_0}&=&H_{w_2w_1}|_{\xi_0}=-\frac{r}{\sigma}\frac{\partial^2U}{\partial w_1\partial w_2}|_{\xi_0}=-\frac{r}{p}\Psi(\bb_{12}).
\end{eqnarray}
Thus the proof is complete.\hb

Theorem \ref{main.theorem.decomposition} immediately follows from Theorem \ref{linearized.Hamiltonian}.

\begin{remark}
If $\beta_{12}=0$, by (\ref{H_12}), we have $H_{w_1w_2}=O$, and hence
the linearized Hamiltonian system (\ref{general.linearized.Hamiltonian.system})
can be separated into three independent Hamiltonian systems,
the first one is the linearized Hamiltonian system of the Kepler two-body problem at Kepler elliptic orbit,
and each of the other two systems can be written as
\be
\dot{\zeta}_i(\th)=JB_{i,0}(\th)\zeta_i(\th), \label{linearized.system.sep_i}
\ee
with
\be
B_{i,0}=\left(\matrix{I_2& -J_2\cr J_2& I_2-{r\over p}\left[{3+\bb_{i,0}\over2}I_2+\Psi(\bb_{ii,0})\right]}\right),
\ee
for $i=1,2$.
Thus the linear stability problem of the elliptic relative equilibrium of the four-body problem
can be reduced to the linear stability problems of system (\ref{linearized.system.sep_i}) with $i=1,2$.
\end{remark}

However in general, $\beta_{12}=0$ does not hold.
But in some special cases, such as the four-body system with two small masses,
we precisely have $\beta_{12}=0$, and we will study such system below.

\section{Appendix: Properties on $\Phi$ and $\Psi$}

Direct computation shows that:
\begin{lemma}
	(i) If $z\in\R$, then
	\be
	\Phi(z)=zI_2,\quad\quad \Psi(z)=z\left(\matrix{1 & 0\cr 0 & -1}\right);
	\ee
	(ii) For any $z\in\C$, we have
	\bea
	\Phi(z)^T=\Phi(\bar{z}),
	\\
	\Psi(z)^T=\Psi(z);
	\eea
	(iii) For any $z,w\in\C$, we have
	\bea
	\Phi(z)\Phi(w)&=&\Phi(zw),
	\\
	\Psi(z)\Psi(w)&=&\Phi(z\bar{w}),
	\\
	\Phi(z)\Psi(w)&=&\Psi(zw),
	\\
	\Psi(z)\Phi(w)&=&\Psi(z\bar{w}).
	\eea
	Specially, we have
	\bea
	\Phi(\bar{z})\Phi(z)=\Phi(z)\Phi(\bar{z})=\Phi(|z|^2)=|z|^2I_2,
	\\
	\Psi(z)\Psi(z)=\Psi(\bar{z})\Psi(\bar{z})=\Phi(|z|^2)=|z|^2I_2.
	\eea
\end{lemma}

\end{document}